\documentclass{PoS}

\title{A model of high-energy emission from jets of microquasars of Population III}

\ShortTitle{Jets of Microquasars of Population III}

\author{\speaker{P. Sotomayor Checa}\\
        Instituto Argentino de Radioastronom\'ia - Consejo Nacional de Investigaciones Cient\'ificas y T\'ecnicas, Argentina\\
        Facultad de Ciencias Astron\'omicas y Geof\'isicas - Universidad Nacional de La Plata, Argentina\\
        E-mail: \email{psotomayor@iar.unlp.edu.ar}}

\author{G.E. Romero\\
        Instituto Argentino de Radioastronom\'ia - Consejo Nacional de Investigaciones Cient\'ificas y T\'ecnicas, Argentina\\
        Facultad de Ciencias Astron\'omicas y Geof\'isicas - Universidad Nacional de La Plata, Argentina\\
        E-mail: \email{gustavo.esteban.romero@gmail.com}}

\abstract{Current research indicates that the radiation of the first stars alone would not have been suficient to ionize the intergalactic medium on long distances because of the high columnar density of the clouds in which they formed. It becomes necessary, then, to explore other alternatives for the reionization. The study of the contribution of the first accreting binary systems to the reionization and heating of the intergalactic medium requires the formulation of a concrete model for the first microquasars. We aim at constructing a consistent model for jets of microquasars where the donor star is of Population III. We calculate the spectral energy distribution of the radiation produced by the relativistic particles in the jets, within the framework of a lepto-hadronic model. In addition, we study the interaction between the jets and the early intergalactic medium. All calculations have been computed for a cosmological redshift value $z=10$. We determine that jets of microquasars of Population III are sources of intense gamma radiation in regions of particle acceleration near the compact object and in the terminal jets.}

\FullConference{International Conference on Black Holes as Cosmic Batteries: UHECRs and Multimessenger Astronomy - BHCB2018\\
		12-15 September, 2018\\
		Foz do Igua\c{c}u, Brasil}

\newcommand{\grad}{$^{\circ}$}

\begin{document}

\section{Introduction}
%%%%%%%%%%%%%%%%%%%%%%%%%%%%%%%%%%%%%%%%%%%%%%%%%%%
%
The primary goal of this work is to start the elaboration of a complete model for jets of microquasars of Population III, in order to provide a tool for quantitative predictions. As a secondary objective we intend to make realistic estimates of the production of radiation and cosmic rays that were injected into the intergalactic medium by these objects. These sources appeared after the gravitational collapse of the first stars. We consider the cosmological redshift value given by $z = 10$. We adopt the physical parameters for Population III binary systems explored in \cite{romero2018}. We assume that the jet launching occurs by a magneto-centrifugal mechanism. In a region close to the compact object, charged particles can be accelerated up to relativistic energies by internal shocks. We adopt a lepto-hadronic model to calculate the radiation produced in the jets by the interaction of the relativistic particles with the different ambient fields. The field of matter that gives rise to the Bremsstrahlung radiation and the inelastic proton-proton collisions is the population of cold protons within the jet. The radiation field for the inverse Compton scattering and the inelastic proton-photon collisions is dominated by the synchrotron radiation field of primary electrons. In addition, we study the production of radiation by the cooling of accelerated electrons in the terminal region of the jet \cite{bordas2009}. \par
%%%%%%%%%%%%%%%%%%%%%%%%%%%%%%%%%%%%%%%%%%%%%%%%%%%%%
\section{Jets}
%%%%%%%%%%%%%%%%%%%%%%%%%%%%%%%%%%%%%%%%%%%%%%%%%%%%%
In a magneto-centrifugal launching mechanism, the jet is ejected by conversion of magnetic energy into kinetic energy. In general, the magnetic field near the compact object has a higher value than equipartition with the matter of the jet. The magnetic field in the jets decreases with the distance z to the compact object. We consider that the flow expands adiabatically, and that:
\begin{equation}
B(z)=B(z_{0})\left(\frac{z_{0}}{z}\right),
\label{equation_MagneticField}
\end{equation}
where $z_{0}$ is the launch point of the jet. The value of $B_{0}=B(z_{0})$ is determined by requiring equipartition between the densities of kinetic and magnetic energy at $z_{0}$:
\begin{equation}
\frac{B^{2}(z_{\mathrm{0}})}{8\pi}=\frac{L_{\mathrm{jet}}}{\pi r_{\mathrm{0}}^{2}v_{\mathrm{jet}}},
\label{equation_EquipartitionDensities}
\end{equation}
where $v_{\mathrm{jet}}$ is the bulk velocity of the jets.\par
The density of cold protons at a distance $z$ from the black hole in the jet is
\begin{equation}
n_{\mathrm{p}}(z) \simeq \frac{\dot{m}_{\mathrm{j}}}{\pi[R_{\mathrm{j}}(z)]^{2}m_{\mathrm{p}}v_{\mathrm{j}}}.
\end{equation}
These cold protons are targets for the relativistic electrons and protons.\par
\begin{figure}[h]
  \centering
  \includegraphics[scale=0.5]{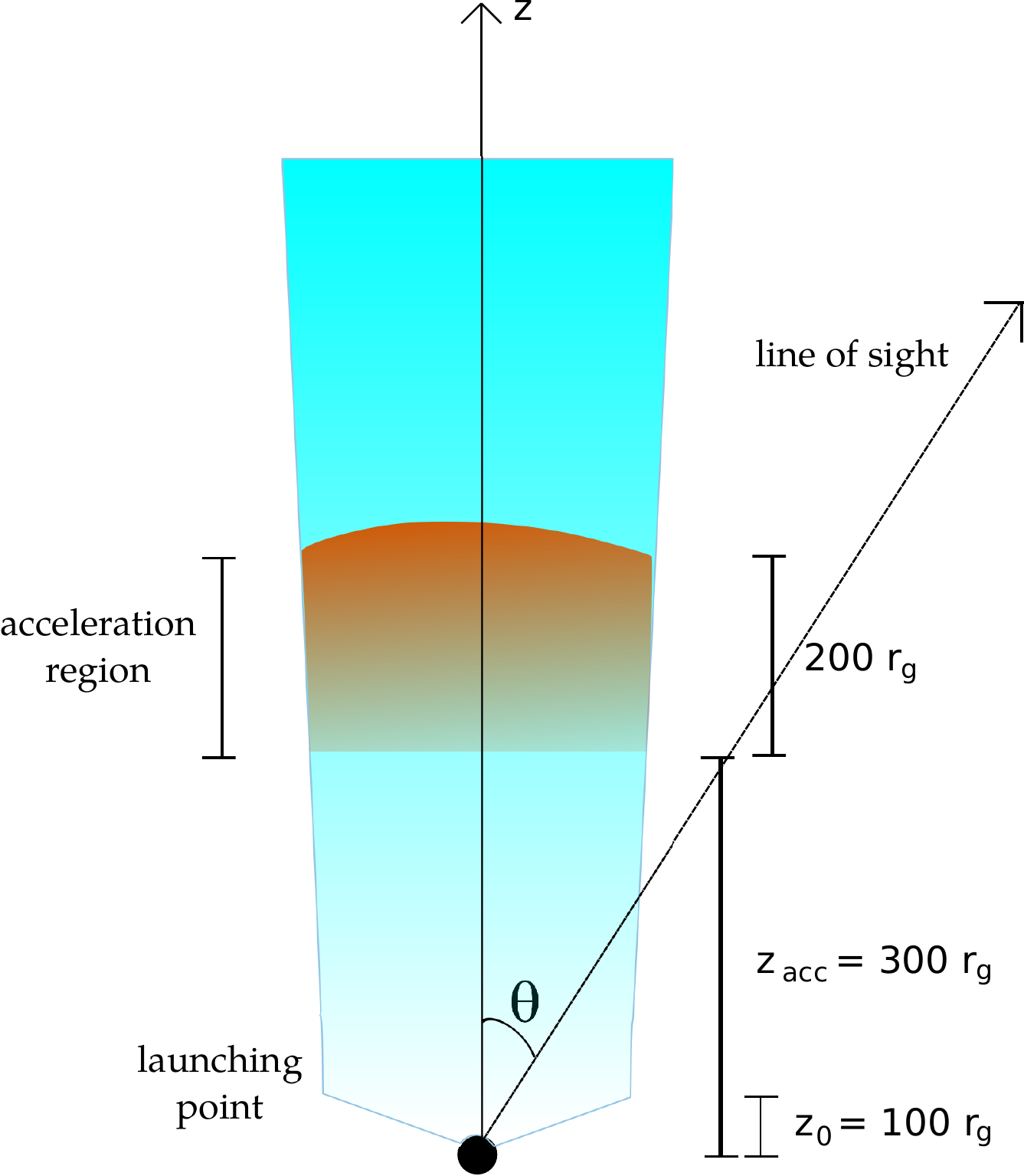}
     \caption{\small Scheme of the jet. It is launched at a distance $z_{0}=100\,r_{\mathrm{g}}$ of the compact object. Acceleration of particles takes place further away at $z_{\mathrm{acc}}=300\,r_{\mathrm{g}}$. The width of the acceleration region is $200\,r_{\mathrm{g}}$. The jet axis makes an angle $\theta$ with
the line of sigh.}
        \label{fig:scheme_jet}
\end{figure}

A sketch of the jet is shown in Fig \ref{fig:scheme_jet}. In the Table \ref{table_JetsParameters} we show the values of the different parameters of the jets.
%%%%%%%%%%%%%%%%%%%%%%%%%%%%%%%%%%%%%%%%%%%%%%%%%%%%%

%%%%%%%%%%%%%%%%%%%%%%%%%%%%%%%%%%%%%%%%%%%%%%%%%%%%%%%%%%%%%%%%%%%%%%%
% Table of jet model parameters
%%%%%%%%%%%%%%%%%%%%%%%%%%%%%%%%%%%%%%%%%%%%%%%%%%%%%%%%%%%%%%%%%%%%%%%
\begin{table*}[h]
\caption{\small Parameters of the jets}             % title of Table
\label{table_JetsParameters}      % is used to refer this table in the text
\centering                          % used for centering table
\begin{tabular}{c c c c}        % centered columns (4 columns)
\hline\hline                 % inserts double horizontal lines
Parameter & Symbol & Value & Unit \\    % table heading 
\hline                        % inserts single horizontal line  
   Jet luminosity & $L_{\mathrm{jet}}$ & $2\times 10^{41}$ & $\mathrm{erg}\,\mathrm{s}^{-1}$ \\
   Jet's bulk Lorentz factor & $\Gamma_{\mathrm{jet}}$ & $9$ & \\
   Jet semi-opening angle tangent & $\chi$ & $0.1$ &  \\
   Jet's content of relativistic particles & $q_{\mathrm{rel}}$ & $0.1$ &  \\ 
   Hadron-to-lepton energy ratio & $a$ & $100$ &  \\
   Jet's launching point & $z_{\mathrm{0}}$ & $100$ & $r_{\mathrm{g}}$ \\
   Magnetic field at base of jet & $B_{\mathrm{0}}$ & $9.8\times 10^{6}$ & $\mathrm{G}$ \\
   Cold matter density inside the jet at $z_{\mathrm{0}}$ & $n_{\mathrm{0}}$ & $5.6\times 10^{14}$ & $\mathrm{cm}^{-3}$ \\
   Injection point & $z_{\mathrm{acc}}$ & $300$ & $r_{\mathrm{g}}$ \\  
   Size of injection point & $\Delta z_{\mathrm{acc}}$ & $200$ & $r_{\mathrm{g}}$ \\  
   Index for magnetic field dependence on $z$ & $m$ & $1.0$ &  \\
   Injection spectral index & $p$ & $2.0$ &  \\
   Acceleration efficiency & $\eta$ & $0.1$ &  \\
   Maximum electron energy & $E_{\mathrm{e}}^{\mathrm{max}}$ & $13.1$ & $\mathrm{GeV}$  \\
   Maximum proton energy & $E_{\mathrm{p}}^{\mathrm{max}}$ & $5.7$ & $\mathrm{PeV}$  \\
\hline                                   %inserts single line
\end{tabular}
\end{table*}
%%%%%%%%%%%%%%%%%%%%%%%%%%%%%%%%%%%%%%%%%%%%%%%%%%%%%%%%%%%%%%%%%%%
% subsubsection: Relativistic particles
%%%%%%%%%%%%%%%%%%%%%%%%%%%%%%%%%%%%%%%%%%%%%%%%%%%%%
\subsection{Relativistic particles}
In the acceleration region a small fraction of the total kinetic power of the jet is transferred into relativistic particles by diffusive acceleration \cite{drury1983}:
\begin{equation}
L_{\mathrm{rel}} = q_{\mathrm{rel}}L_{\mathrm{jet}},
\label{equation_LuminosityRel}
\end{equation} 
where we adopt $q_{\mathrm{rel}} = 0.1$ and $L_{\mathrm{rel}}$ is the total power injected into both relativistic protons and electrons, $L_{\mathrm{rel}} = L_{\mathrm{p}} + L_{\mathrm{e}}$.
In addition, we assume that the jet is proton-dominated (see e.g. \cite{romero2008}):
\begin{equation}
L_{\mathrm{p}} = aL_{\mathrm{e}} \qquad a=100.
\label{equation_LuminosityDiv}
\end{equation} 
The protons and electrons that are injected by the mechanism that accelerates thermal particles to relativistic energies are called primary particles. The secondary particles are the pions, muons and electron-positron pairs injected as a result of the interaction of primary particles with matter and radiation. Injection and cooling of primary and secondary particles occur in the acceleration region. We assume an injection function that is a power-law in the energy of the primary particles and that is inversely proportional to the distance to the compact object \cite{romero2008}:
\begin{equation}
Q^{\prime}(E^{\prime},z^{\prime}) = Q_{\mathrm{0}}\frac{{E^{\prime}}^{-p}}{z^{\prime}} \qquad \left[Q^{\prime}\right] = \mathrm{erg}^{-1}\mathrm{s}^{-1}\mathrm{cm}^{-3},
\label{equation_InyectionParticles}
\end{equation}
where $p$ is the spectral index of particle injection. In this proceeding, the primed variables are measured in the co-moving reference frame of the jet, while unprimed variables in the observer frame. The normalization constant $Q_{\mathrm{0}}$ is obtained as:
\begin{equation}
L_{(\mathrm{e,p})} = \int _{V} \mathrm{d}^{3}r \int ^{E_{(\mathrm{e,p})} ^{\mathrm{max}}} _{E_{(\mathrm{e,p})}^{\mathrm{min}}}\mathrm{d}E_{(\mathrm{e,p})}E_{(\mathrm{e,p})}Q_{(\mathrm{e,p})}(E_{(\mathrm{e,p})},z),
\label{equation_LuminosityParticles}
\end{equation}
where $V$ is the volume of the acceleration region. The maximum energy $E^{\mathrm{max}}$ that a relativistic particle can attain is obtained by balancing its rate of acceleration and cooling.\par 
The acceleration rate for a charged particle is:
\begin{equation}
{t^{\prime}}^{-1}_{\mathrm{acc}} = \frac{\eta e c B^{\prime}}{E^{\prime}},
\label{equation_AccelerationRate}
\end{equation}
where $B^{\prime}$ is the magnetic field, $E^{\prime}$ is the energy of the particle, and $\eta$ is a parameter that characterizes the efficiency of the acceleration mechanism.\par
The cooling rate is the sum of the radiative cooling rate, adiabatic losses rate, and escape rate:
\begin{equation}
{t^{\prime}}^{-1}_{\mathrm{cool}} = {t^{\prime}}^{-1}_{\mathrm{rad}} + {t^{\prime}}^{-1}_{\mathrm{ad}} + {t^{\prime}}^{-1}_{\mathrm{esc}}.
\label{equation_CoolingRate}
\end{equation}
The radiative losses are caused by the interaction of the relativistic particles with the fields in the jet. We consider electron losses by synchrotron radiation, inverse Compton scattering, and relativistic Bremsstrahlung, and proton losses by synchrotron radiation, inelastic proton-proton collisions, and photo-hadronic interactions.
%%%%%%%%%%%%%%%%%%%%%%%%%%%%%%%%%%%%%%%%%%%%%%%%%%%%%
% Total radiative losses
%%%%%%%%%%%%%%%%%%%%%%%%%%%%%%%%%%%%%%%%%%%%%%%%%%%%%
Total radiative losses for relativistic electrons are calculated as:
\begin{equation}
{t^{\prime}}^{-1}_{\mathrm{rad}} = {t^{\prime}}^{-1}_{\mathrm{synchr}} + {t^{\prime}}^{-1}_{\mathrm{IC}} + {t^{\prime}}^{-1}_{\mathrm{Br}},
\end{equation}
whereas for relativistic protons they are:
\begin{equation}
{t^{\prime}}^{-1}_{\mathrm{rad}} = {t^{\prime}}^{-1}_{\mathrm{synchr}} + {t^{\prime}}^{-1}_{p\gamma, e^{\pm}} + {t^{\prime}}^{-1}_{p\gamma, \pi} + {t^{\prime}}^{-1}_{pp}.
\end{equation}
%%%%%%%%%%%%%%%%%%%%%%%%%%%%%%%%%%%%%%%%%%%%%%%%%%%%%
\section{Radiative processes}
We calculate the spectral energy distribution of the radiation produced by primary and secondary particles in their interaction with magnetic, matter and radiation fields within the jet. Detailed formulas for the photon emissivity of each radiative process can be found in \cite{romero2008,bosch-ramon2006,reynoso2009,
romero2014,vilatesis}.\par
Particle distributions obtained by solving the transport equation are valid in the reference frame of the jet, where the particle distributions are isotropic. The luminosities are first obtained in this reference frame, and then transformed to the observer frame. Since we consider a conical jet, if $q_{\gamma}^{\prime}$ is the photon emissivity (in $\mathrm{erg\,s^{-1}\,cm^{-3}}$) in the reference frame of the jet associated with some radiative process, the luminosity at energy $E_{\gamma}^{\prime}$ is:
\begin{equation}
L_{\gamma}^{\prime} (E_{\gamma}^{\prime}) = \pi \int_{z^{\prime}_{\mathrm{acc}}}^{z^{\prime}_{\mathrm{max}}} \mathrm{d}z^{\prime} \, {r^{\prime}}^{2}_{\mathrm{jet}}(z^{\prime}) \, q_{\gamma}^{\prime}(E_{\gamma}^{\prime},z^{\prime}).
\label{equation_LuminosityJetFrame}
\end{equation}
The photon energy in the observer frame is obtained by applying the appropiate boost:
\begin{equation}
E_{\gamma} = D\,E_{\gamma}^{\prime}.
\end{equation}
where $D$ is the factor Doppler. For an approaching jet with viewing angle $\theta$,
\begin{equation}
D = \frac{1}{\Gamma_{\mathrm{jet}}\left(1 - \beta_{\mathrm{jet}}\cos \theta\right)}.
\end{equation}
The luminosity in the observer frame is:
\begin{equation}
L_{\gamma}(E_{\gamma}) = \frac{D^{2}}{\Gamma_{\mathrm{jet}}} L_{\gamma}^{\prime} (E_{\gamma}^{\prime}).
\end{equation}
In the case of $pp$ inelastic collisions the luminosity is conveniently calculated in the reference frame of the jet, where some useful parameterizations for the cross section are available. For this, it is necessary to convert the particle distribution obtained to the rest frame of the observer \cite{torres2011}.\par
%%%%%%%%%%%%%%%%%%%%%%%%%%%%%%%%%%%%%%%%%%%%%%%%%%%%%
% SED of the radiation produced by primary particles
%%%%%%%%%%%%%%%%%%%%%%%%%%%%%%%%%%%%%%%%%%%%%%%%%%%%%
\begin{figure}[h!]
 \centering
 \includegraphics[width=0.6\hsize]{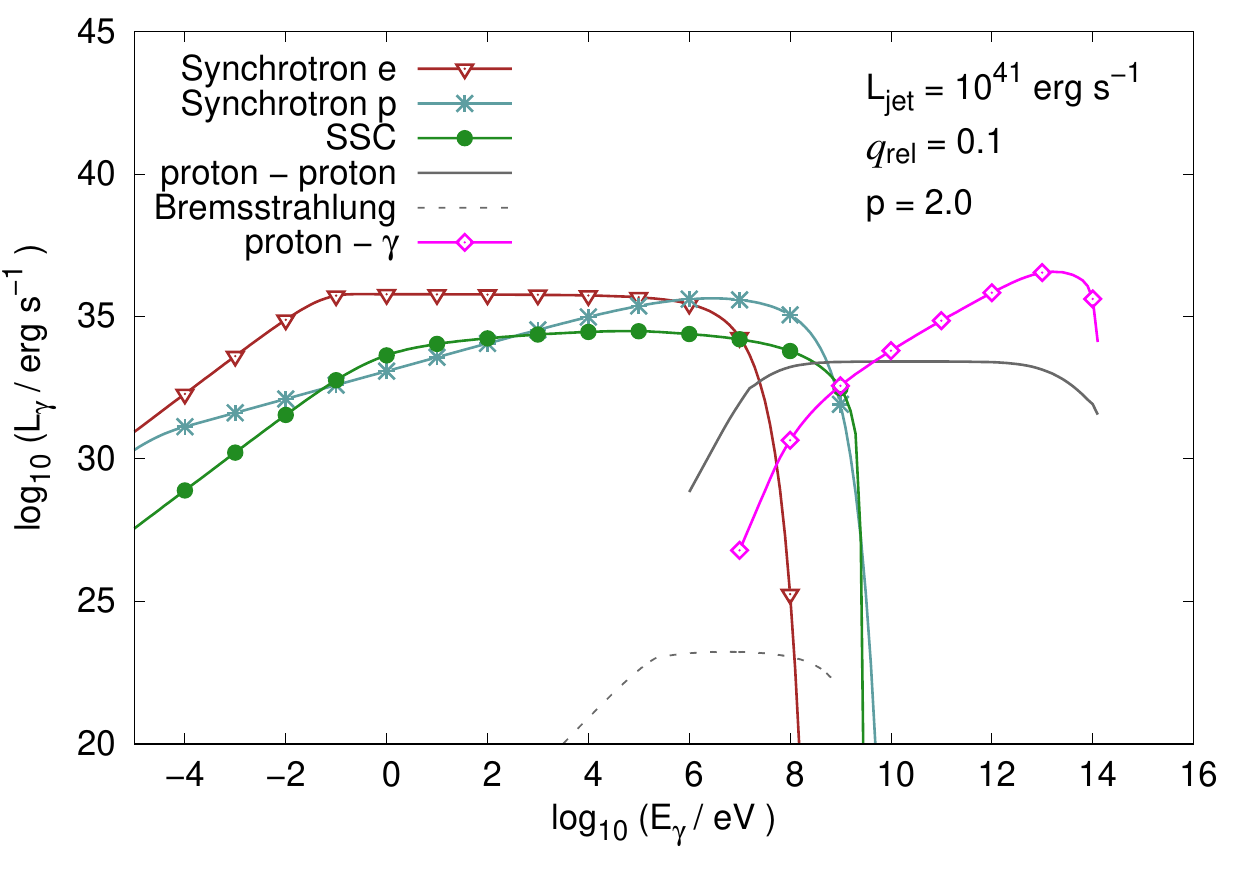}
    \caption{\small Non-thermal spectral energy distribution of the radiation produced in the jet by the population of primary particles. The adopted viewing angle is $\theta$ = 60\grad.}
       \label{fig:SEDtotalPrimaries}
 \end{figure}
%%%%%%%%%%%%%%%%%%%%%%%%%%%%%%%%%%%%%%%%%%%%%%%%%%%%
Figure \ref{fig:SEDtotalPrimaries} show the SEDs of the non-thermal radiation produced in the jet by the population of primary particles. The interactions of the relativistic electrons and protons with the fields in the jet give rise to radiation in almost all the electromagnetic spectrum. For energies lower than 1 MeV, the electromagnetic emission is mainly caused by the synchrotron radiation of electrons. For energies in the range between 1 MeV and 1 GeV, it is dominated by the synchrotron radiation of protons. In the range between 1 GeV and 10 GeV the radiation is mainly due to inelastic collisions $pp$. Finally, for energies greater than 10 GeV it is mainly the result of the $pp$ and $p\gamma$ collisions. The maximum value of the emitted luminosity is from the $p\gamma$ collisions, $L_{p\gamma}\approx 10^{37}\,\mathrm{erg\,s^{-1}}$. The contribution by relativistic Bremsstrahlung is irrelevant. The SEDs are corrected by Doppler effect considering a viewing angle of $\theta$ = 60\grad.
%%%%%%%%%%%%%%%%%%%%%%%%%%%%%%%%%%%%%%%%%%%%%%%%%%%%%
\subsection{Secondary pair injection}
Three channels for secondary pair production were taken into account: pair production as a result of muon decay after $pp$ and $p\gamma$ interactions, pairs produced directly in $p\gamma$ interactions (Bethe-Heitler pairs), and pairs created in photon-photon annihilation. In order to calculate the contribution of the radiation produced by the secondary pairs to the total luminosity, the injection function of the pairs must be calculated. \par
For $p\gamma$ interactions, the injection function of the pairs can be estimated by the approximation developed in \cite{atoyan2003}. For $pp$ inelastic colissions, we have calculated the injection function of the pairs from the injection function of charged pions, as it is developed in \cite{kelner2006}.
Bethe-Heitler pairs are produced directly in $p\gamma$ interactions when the photon energy is higher to $1.022\,\mathrm{MeV}$ in the rest frame of the proton. In the approximation of the $\delta$-functional, the injection function of pairs is given by \cite{mastichiadis2005}.\par
The last mechanism of pair creation that we consider is $\gamma \gamma$ annihilation. The target photon field that we consider is the synchrotron radiation field of primary electrons. The injection function can be approximated by \cite{aharonian1983}.\par
%%%%%%%%%%%%%%%%%%%%%%%%%%%%%%%%%%%%%%%%%%%%%%%%%%%%%
% SED of the synchrotron radiation produced by electron-positron pairs.
%%%%%%%%%%%%%%%%%%%%%%%%%%%%%%%%%%%%%%%%%%%%%%%%%%%%%
\begin{figure}[h!]
 \centering
 \includegraphics[width=0.6\hsize]{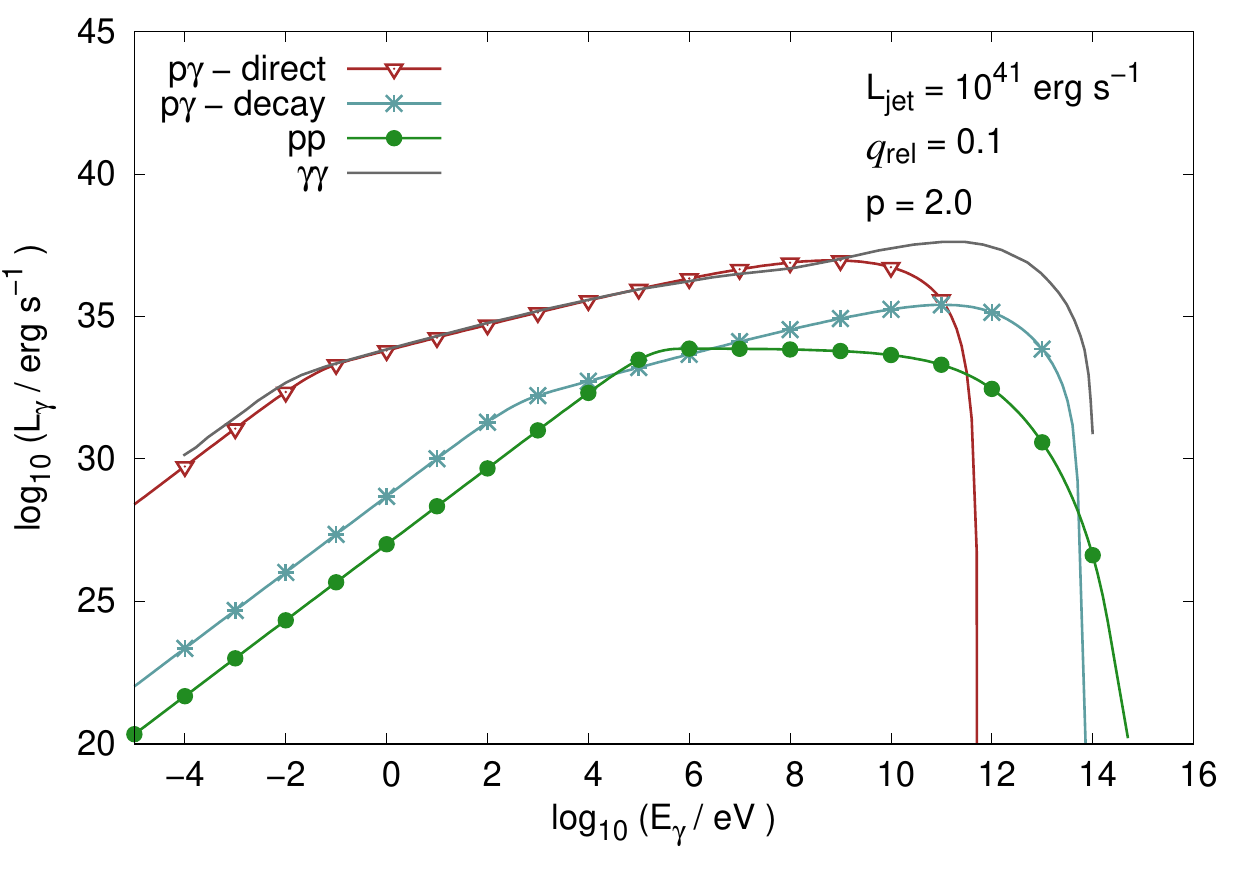}
    \caption{\small Spectral energy distribution of the synchrotron radiation for electron-positron pairs. The adopted viewing angle is $\theta$ = 60\grad.}
       \label{fig:SED_pairs}
 \end{figure}
\begin{figure}[h!]
 \centering
 \includegraphics[width=0.6\hsize]{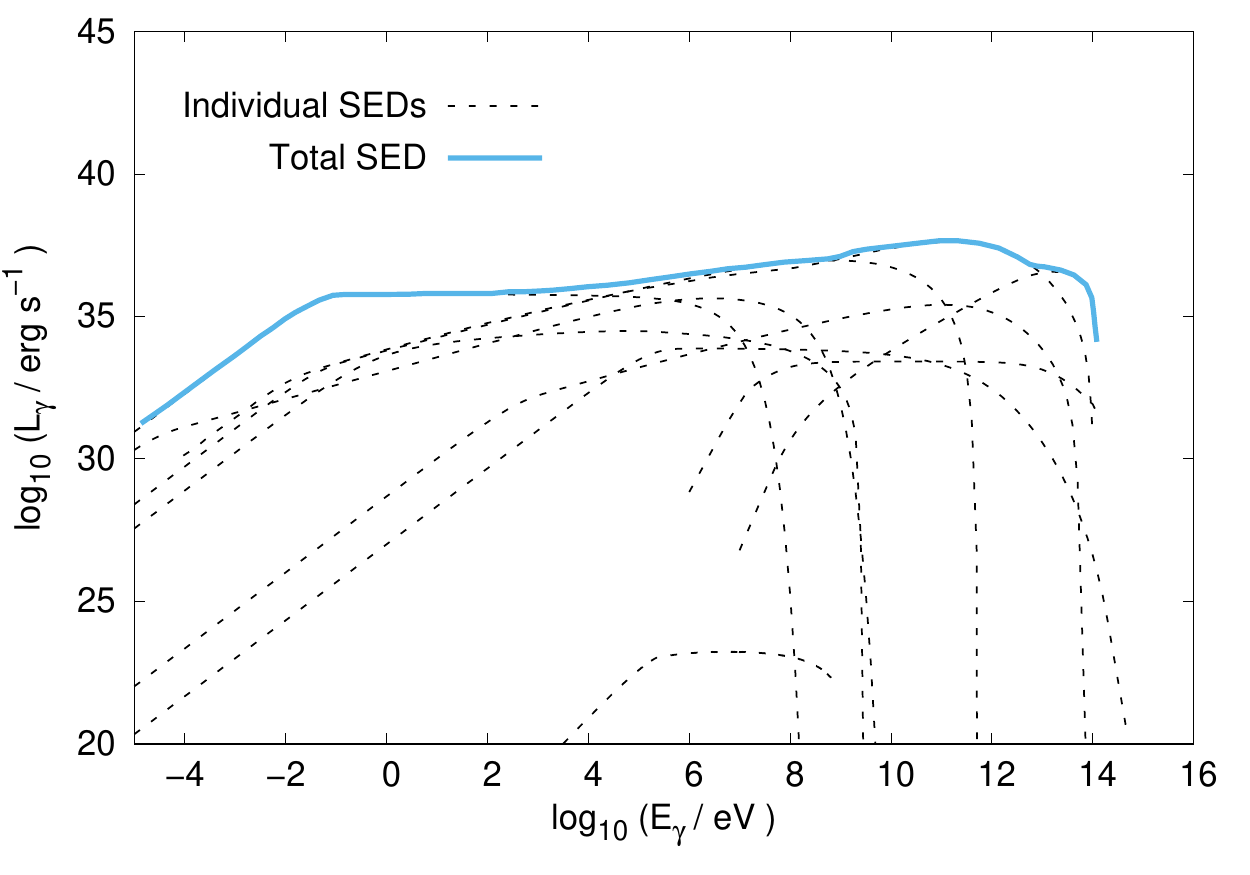}
    \caption{\small Total non-thermal spectral energy distribution of the jet without attenuation. Dotted lines show individual contributions. The adopted viewing angle is $\theta$ = 60\grad.}
       \label{fig:SED_total_jet}
 \end{figure}

%%%%%%%%%%%%%%%%%%%%%%%%%%%%%%%%%%%%%%%%%%%%%%%%%%%%%
The secondary pairs produced in the jets are cooled mainly by synchrotron radiation. The synchrotron spectral energy distributions of the secondary particles are shown in Fig. \ref{fig:SED_pairs}. These SEDs must be added to the SEDs produced by the primary particles for the calculation of the total SED, which is shown in Fig. \ref{fig:SED_total_jet}. It is observed that the emission with greatest luminosity corresponds to the synchrotron radiation of the pairs by $\gamma \gamma $ annihilation. This is expected since all photons emitted with $E_{\gamma} > 10\,\mathrm{MeV}$ are absorbed.
\par
%%%%%%%%%%%%%%%%%%%%%%%%%%%%%%%%%%%%%%%%%%%%%%%%%%%%%
% SED total of the jet.
%%%%%%%%%%%%%%%%%%%%%%%%%%%%%%%%%%%%%%%%%%%%%%%%%%%%%
%%%%%%%%%%%%%%%%%%%%%%%%%%%%%%%%%%%%%%%%%%%%%%%%%%%%%
\subsection{Absorption of the radiation}
\begin{figure}[h]
  \centering
  \includegraphics[width=0.55\hsize]{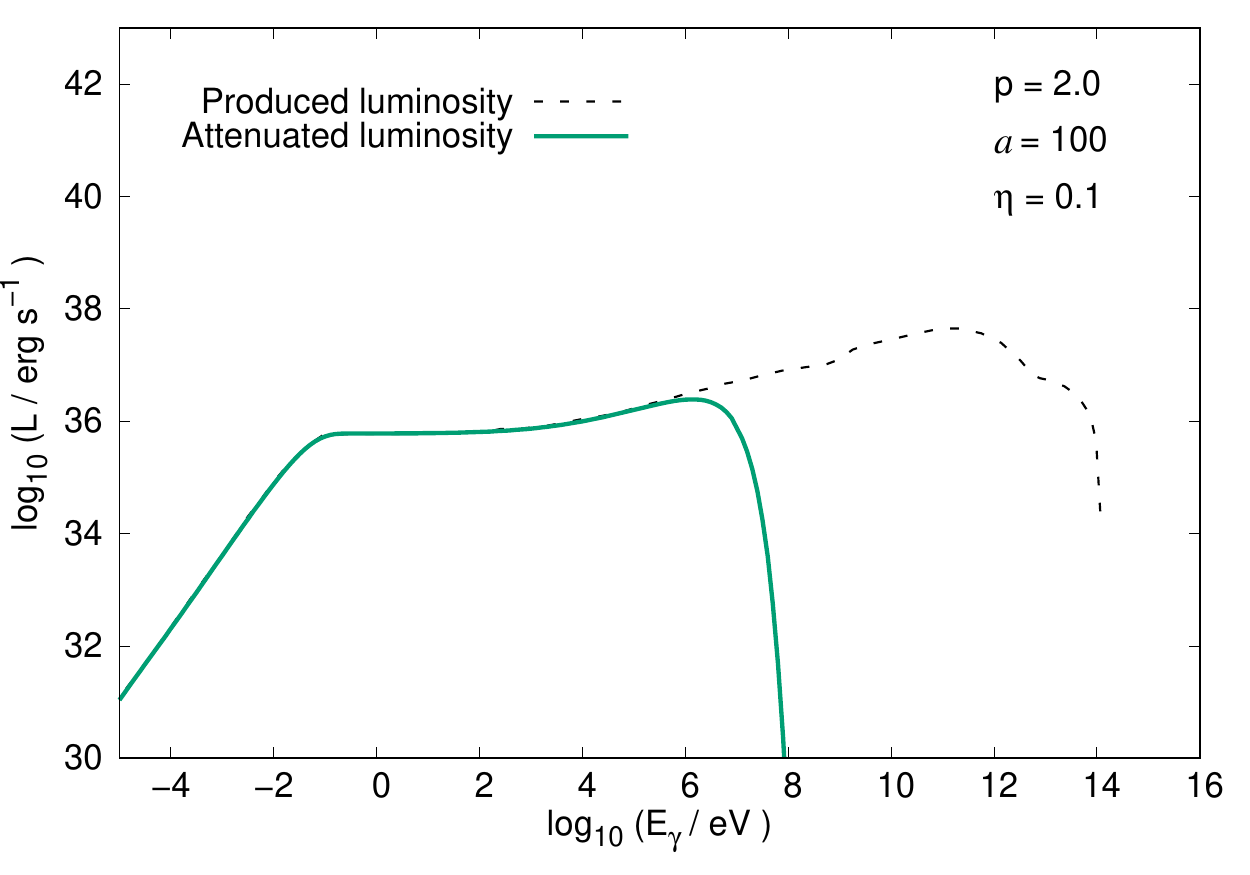}
     \caption{\small Total non-thermal spectral energy distribution of the jet attenuated by internal absorption.}
        \label{SED_total_attenuated}
  \end{figure}

%%%%%%%%%%%%%%%%%%%%%%%%%%%%%%%%%%%%%%%%%%%%%%%%%%%%%
Absorption by $\gamma \gamma$ annihilation with synchrotron photons of the primary electrons produced internally attenuates the emission of the jets for energies greater than 10 MeV, as shown in Fig. \ref{SED_total_attenuated}. This absorbed radiation produces electron-positron pairs which are created with very high energies and are cooled mainly by synchrotron radiation, whereby the energy of the initial photons is distributed in synchrotron photons of lower energy, with no electromagnetic cascades taking place, since they are suppressed by the high magnetic fields. Absorption by $\gamma \gamma$ annihilation with the radiation field of the star can be considered. However, this mechanism only partially attenuates the emitted radiation, being irrelevant in comparison with the internal absorption.\par
%%%%%%%%%%%%%%%%%%%%%%%%%%%%%%%%%%%%%%%%%%%%%%%%%%%%%
% Corrected SED by internal absorption
%%%%%%%%%%%%%%%%%%%%%%%%%%%%%%%%%%%%%%%%%%%%%%%%%%%%%
%%%%%%%%%%%%%%%%%%%%%%%%%%%%%%%%%%%%%%%%%%%%%%%%%%%%%
%%%%%%%%%%%%%%%%%%%%%%%%%%%%%%%%%%%%%%%%%%%%%%%%%%%%
\section{Emission from the terminal jets}
\begin{figure}[h!]
  \centering
  \includegraphics[scale=0.4]{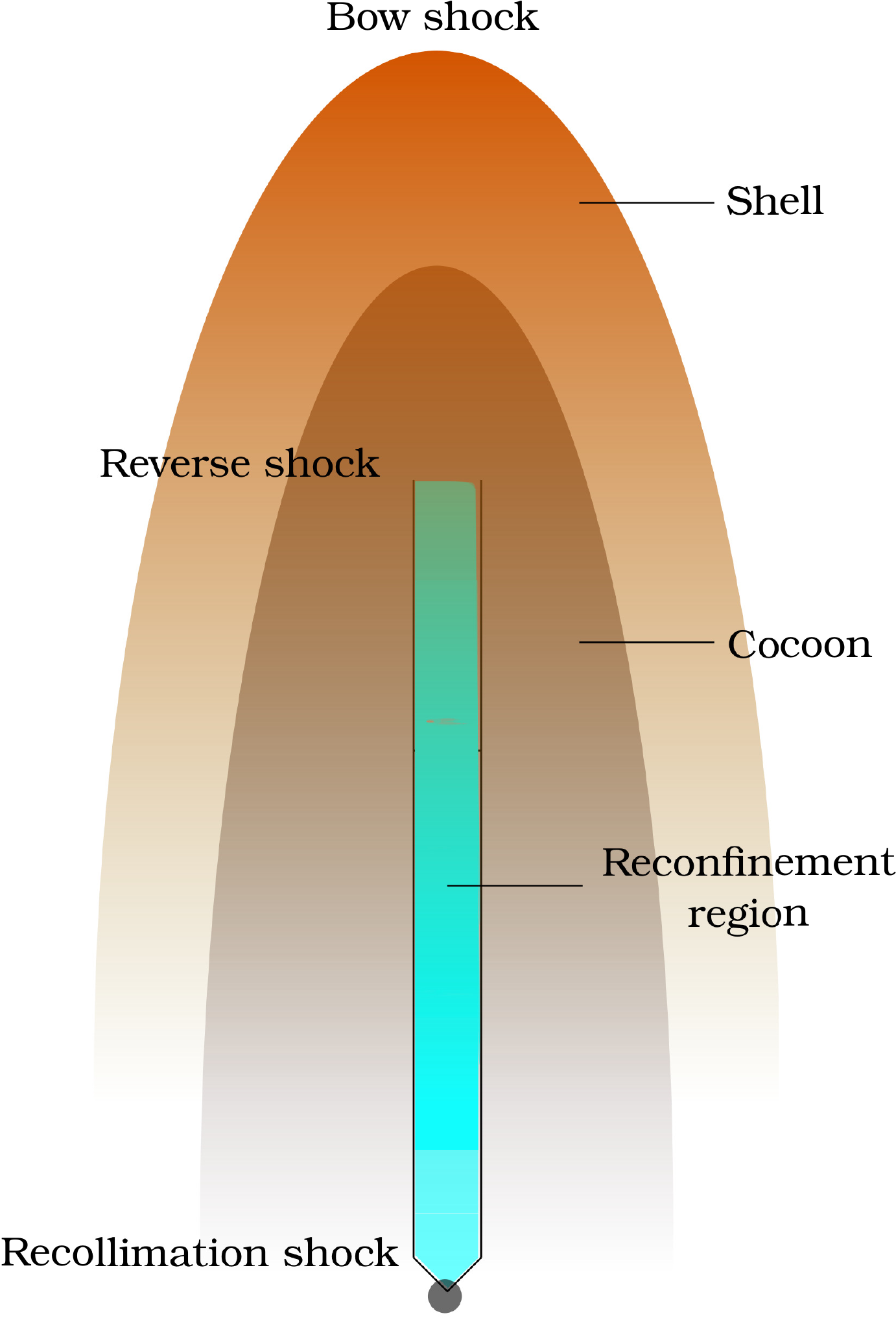}
     \caption{\small Scheme of the terminal region of the jet.}
        \label{fig:scheme_terminaljet}
\end{figure}

%%%%%%%%%%%%%%%%%%%%%%%%%%%%%%%%%%%%%%%%%%%%%%%%%%%%
The jet begins to decelerate when the mass swept in the medium equals the mass transported by the jet. Then, two schocks form in the head of the jet: a bow-shock propagating in the interstellar medium and a reverse-shock directed towards the interior of the jet. The matter that crosses the reverse shock inflates a region of the jet known as the cocoon. At the point of the jet where the pressure is equal to the pressure of the cocoon, another shock called recollimation-shock is formed. A sketch of the system is shown in Fig. \ref{fig:scheme_terminaljet} .\par
%%%%%%%%%%%%%%%%%%%%%%%%%%%%%%%%%%%%%%%%%
%%%%%%%%%%%%%%%%%%%%%%%%%%%%%%%%%%%%%%%%%%%%%%
We do an elementary analysis of these three regions, where particle acceleration can take place, considering two different epochs for the microquasar age $t_{\mathrm{MQ},1} = 10^{4}\,\mathrm{yrs}$ and $t_{\mathrm{MQ},2} = 10^{5}\,\mathrm{yrs}$. For the calculations we apply the model developed by \cite{bordas2009}.\par 
%%%%%%%%%%%%%%%%%%%%%%%%%%%%%%%%%%%%%%%%%%%%%%%%%%%%%
\begin{table}[h!]
\label{table:JetTerminal}      % is used to refer this table in the text
\centering                          % used for centering table
\begin{tabular}{c c c c c c}        % centered columns (4 columns)
\hline\hline                 % inserts double horizontal lines
Parameter & Symbol &  & Value &  & Unit \\    % table heading 
\hline                        % inserts single horizontal line
   Jet luminosity & $L_{\mathrm{jet}}$ & & $10^{41}$ & & $\mathrm{erg}\,\mathrm{s}^{-1}$ \\
   Intergalactic medium density & $n_{\mathrm{IGM}}$ & & $2.4\times 10^{-4}$ & & $\mathrm{cm}^{-3}$ \\
   Source age & $t_{\mathrm{MQ}}$ & $10^{4}$ & & $10^{5}$ & yrs \\
\hline
   \textit{SHELL} \\
\hline                        % inserts single horizontal line
   Magnetic field & $B$ & $5.7\times 10^{-5}$ & & $2.3\times 10^{-5}$ & $\mathrm{G}$ \\
   Shock velocity & $v_{\mathrm{b}}$ & $7.2\times 10^{8}$ & & $2.9\times 10^{8}$ & $\mathrm{cm}\,\mathrm{s}^{-1}$ \\
   Emitter size & $r$ & $1.3\times 10^{20}$ & & $5\times 10^{20}$ & $\mathrm{cm}$ \\
   Maximum energy & $E_{\mathrm{max}}$ & $3.1$ & & $3.3$ & $\mathrm{TeV}$  \\
   Target density & $n_{\mathrm{t}}$ & $9.6\times 10^{-4}$ & & $9.6\times 10^{-4}$ & $\mathrm{cm}^{-3}$ \\
\hline    
   \textit{COCOON} \\
\hline                        % inserts single horizontal line
   Magnetic field & $B$ & $4.6\times 10^{-4}$ & & $1.9\times 10^{-4}$ & $\mathrm{G}$ \\
   Shock velocity & $v_{\mathrm{sh}}$ & $2.98\times 10^{10}$ & & $2.98\times 10^{10}$ & $\mathrm{cm}\,\mathrm{s}^{-1}$ \\
   Emitter size & $r$ & $2.5\times 10^{18}$ & & $6.3\times 10^{18}$ & $\mathrm{cm}$ \\
   Maximum energy & $E_{\mathrm{max}}$ & $1.0$ & & $1.0$ & $\mathrm{PeV}$  \\
\hline  
   \textit{RECONFINEMENT} \\
\hline                        % inserts single horizontal line   
   Magnetic field & $B$ & $4.6\times 10^{-3}$ & & $1.9\times 10^{-3}$ & $\mathrm{G}$ \\
   Shock velocity & $v_{\mathrm{conf}}$ & $2.98\times 10^{9}$ & & $2.98\times 10^{9}$ & $\mathrm{cm}\,\mathrm{s}^{-1}$ \\
   Emitter size & $r$ & $3.8\times 10^{20}$ & & $1.5\times 10^{21}$ & $\mathrm{cm}$ \\
   Maximum energy & $E_{\mathrm{max}}$ & $33.4$ & & $33.4$ & $\mathrm{TeV}$  \\
\hline    
\end{tabular}
\caption{\small Parameter values adopted for the three emitting zones in the jet's terminal region}             % title of Table
\end{table}
%%%%%%%%%%%%%%%%%%%%%%%%%%%%%%%%%%%%%%%%%%%%%%%%%%%%%
We assume that the magnetic field in the downstream regions is such that the magnetic energy density is $\sim10\%$ of the thermal energy density of the gas. In addition, in each acceleration region, the fraction of kinetic power converted into non-thermal particles is set at $1\%$ of the power of the jet. We have analyzed only the leptonic contribution here.\par
The injection function in the recollimation-shock, the reverse-shock, and the bow-shock, is assumed to be by $Q(E) = Q_{0}\, E^{-2}$, with the normalization constant $Q_{0}$ calculated as in Eq. (\ref{equation_LuminosityParticles}).\par
The cooling processes for non-thermal electrons in these regions are: synchrotron radiation, inverse Compton scattering, and adiabatic losses. The target radiation field for inverse Compton losses is the cosmic microwave background ($T_{\mathrm{CMB}} = 30\,\mathrm{K}$, for $z=10$). In the reconfinement region there are not adiabatic losses since the jet radius keeps constant. Because of the low density of matter in the cocoon and the reconfinement region, relativistic Bremsstrahlung radiation is negligible in these two zones and is only computed in the shell. \par
The model parameters for the three emitting zones in the terminal region of the jets that we adopt are listed in Table \ref{table:JetTerminal}.\par
%%%%%%%%%%%%%%%%%%%%%%%%%%%%%%%%%%%%%%%%%%%%%%%%%%%%%
% SED of the radiation produced in the terminal regions of the jet.
%%%%%%%%%%%%%%%%%%%%%%%%%%%%%%%%%%%%%%%%%%%%%%%%%%%%%
\begin{figure}[h!]
\centering
  \begin{minipage}{0.6\textwidth}
    \includegraphics[width=0.8\textwidth]{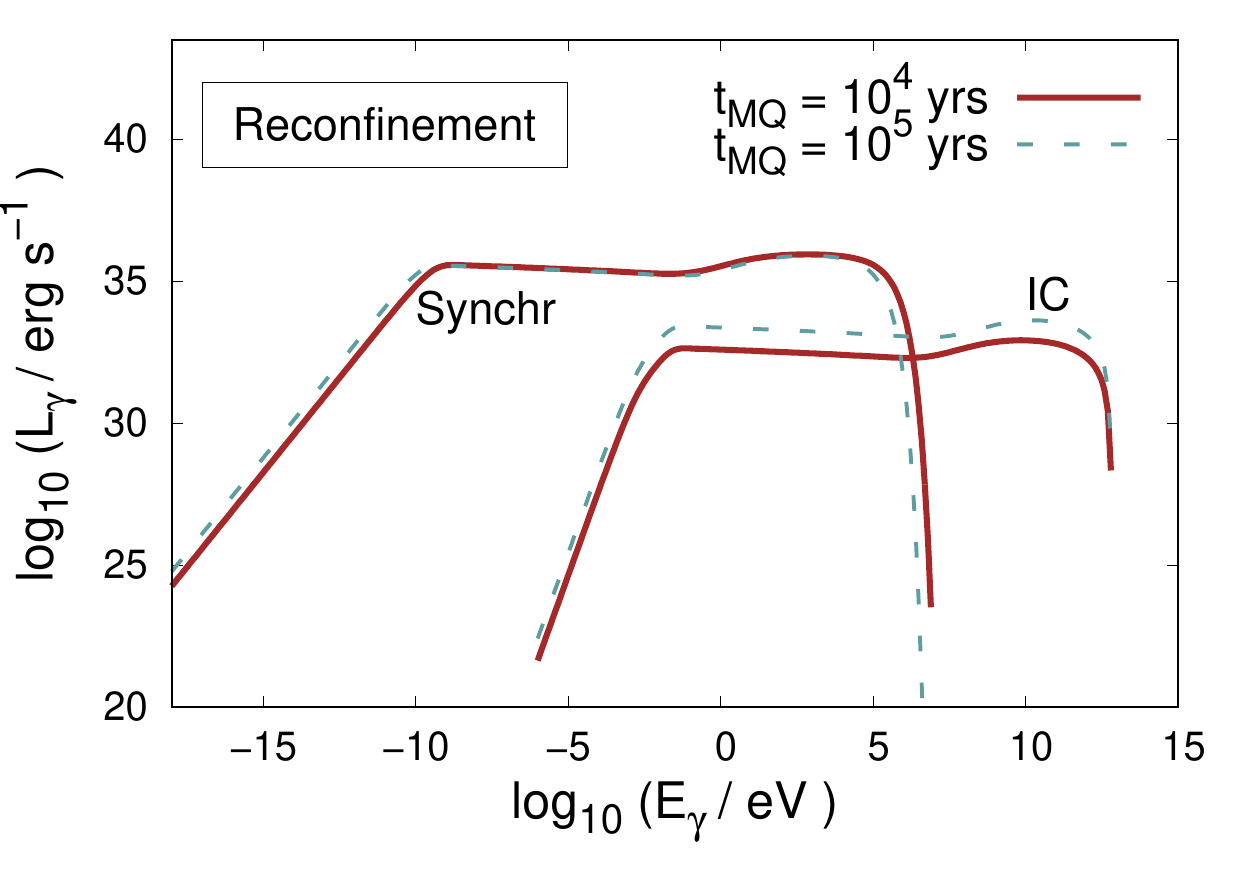}\\
   \end{minipage}%
  \hspace{5mm}
  \begin{minipage}{0.6\textwidth}
    \includegraphics[width=0.8\textwidth]{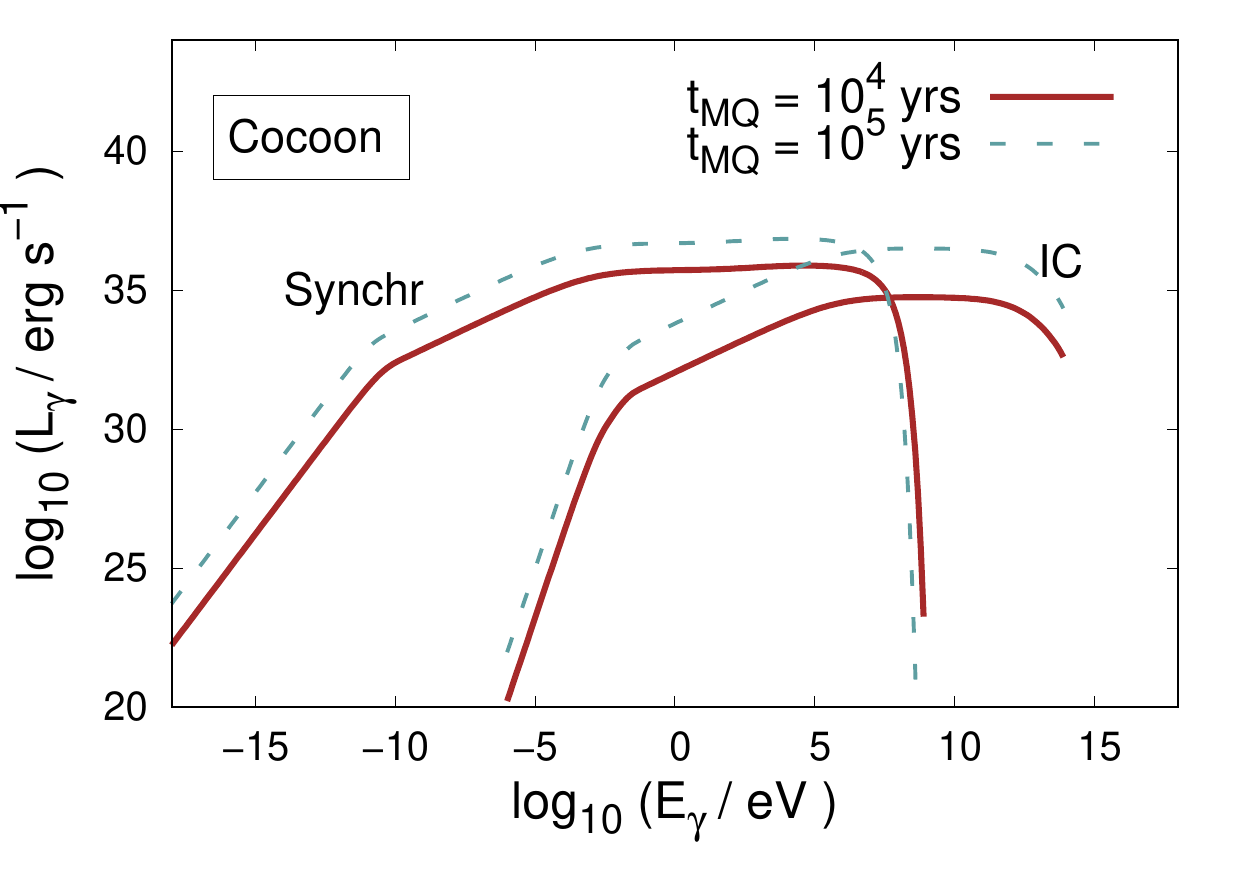}\\
  \end{minipage}
  \hspace{5mm}
  \begin{minipage}{0.6\textwidth}
    \includegraphics[width=0.8\textwidth]{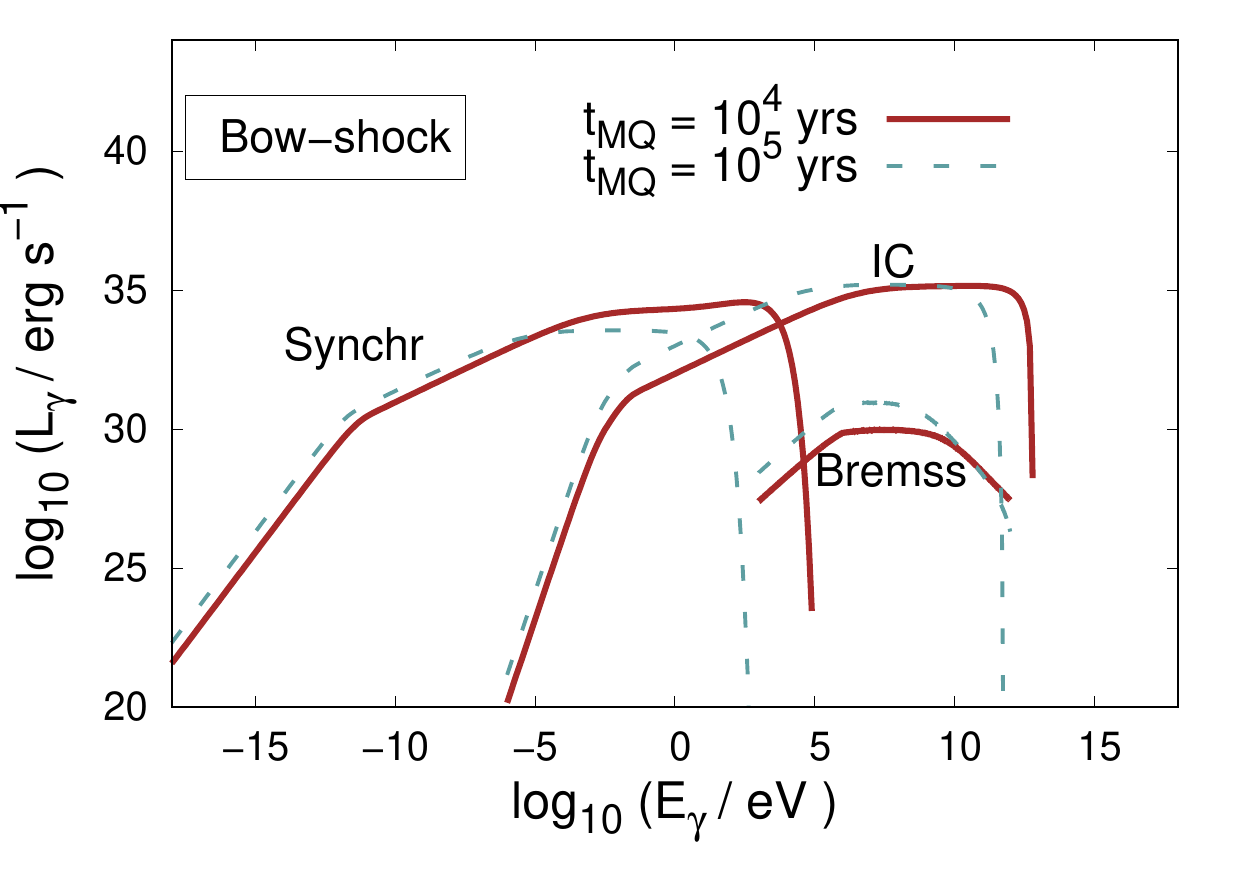}\\
  \end{minipage} 
  \caption{\small Non-thermal spectral energy distributions of the radiation produced in the reconfinement region, cocoon and bow-shock. Two epochs are considered for the source.}
  \label{fig:SED_terminaljet}
\end{figure}
%%%%%%%%%%%%%%%%%%%%%%%%%%%%%%%%%%%%%%%%%%%%%%%%%%%%%
In Fig. \ref{fig:SED_terminaljet} we show the spectral energy distribution of the radiation produced in the reconfinement region, cocoon, and the shell by the relativistic electrons, for the two epochs of interest. In the region of reconfinement, the maximum energies of electrons for the different epochs are approximately 30 TeV, being always limited by synchrotron losses. Synchrotron emission is also the dominant radiative process. In the cocoon, the shape of the non-thermal spectral energy distribution differs from that seen in the reconfinement region. This is because in the cocoon the cooling mechanism of the electrons varies according to the energy they have: for energies less than $10\,\mathrm{GeV}$ the main cooling mechanism is adiabatic losses, for energies between $10\, \mathrm{GeV}$ and $10\,\mathrm{TeV}$ the main mechanism is inverse Compton scattering, finally for larger energies the mechanism is synchrotron radiation. On the contrary, in the reconfinement region, the cooling mechanism in the whole range of energies is synchrotron. In the cocoon, the two contributions to the bolometric luminosity are comparable, in both epochs of the source. In the shell, the luminosity produced by inverse Compton scattering dominates the emission.  We have included the spectral energy distribution by relativistic Bremsstrahlung radiation, which turns out to be irrelevant because of the low density of the IGM to high redshift. In the same sense, the radiation produced by $pp$ collisions will be irrelevant. So, a leptonic model is appropriate to describe the proposed scenario. This differs from the results obtained in known microquasars, where the emission by relativistic Bremsstrahlung could dominate the spectrum at high energies \cite{bordas2009}.
\par
\section{Discussion}
%%%%%%%%%%%%%%%%%%%%%%%%%%%%%%%%%%%%%%%%%%%%%%%%%%%%%
From the results showed in the previous section we conclude that jets in microquasars of Population III can be significant $\gamma$-rays sources in the range MeV - GeV because of internal shocks in regions near to compact object. Emission from the jets in the range GeV - PeV should be produced in the terminal region where the internal and external absorption is negligible. Similarly, accelerated electrons in the terminal region of the jets can reach very-high energies, because the cooling mechanism by synchrotron radiation becomes inefficient.\par
We have considered a lepto-hadronic jet model in the near region to the compact object, where the power of relativistic particles is dominated by protons. It is easy to calculate lepton-dominated jet models. In such models it is expected a greater absorption caused by $\gamma \gamma$ annihilation with the synchrotron radiation field of primary electrons \cite{romero2008}.\par
Kelvin-Helmholtz instabilities in the contact surface between the wind of the accretion disk and the jet are expected. This will be treated in a future work. In addition, other launch and collimation mechanisms of the jets should be considered. An interesting proposal in this sense that will be explored is the model of magnetically collimated and radiation-pressure driven jet (RMHD jet) developed by \cite{takeuchi2010}. The RMHD jet is accelerated by the radiation-pressure and is collimated by the Lorentz force of a magnetic tower.\par
%%%%%%%%%%%%%%%%%%%%%%%%%%%%%%%%%%%%%%%%%%%%%%%%%%%%
\section{Conclussions}
%%%%%%%%%%%%%%%%%%%%%%%%%%%%%%%%%%%%%%%%%%%%%%%%%%%%
We have developed a simple model for jets of microquasars where the donor star is from Population III  placing special emphasis on the high-energy emission from these hypothetical sources. We have considered a proton-dominated microquasar that predicts significant gamma-ray emission and production of high-energy cosmic rays in the terminal regions of the jets. Within the jets, in the region of particle acceleration near the compact object, the effects of absorption by internal $\gamma\gamma$ annihilation are dramatic: radiation is completely suppressed for energies greater than MeV.\par
%%%%%%%%%%%%%%%%%%%%%%%%%%%%%%%%%%%%%%%%%%%%%%%%%%


\begin{thebibliography}{99}

\bibitem{romero2018} G.E. Romero, \& P. Sotomayor Checa,
\emph{Population III Microquasars},
\emph{IJMPD} {\bf 27} (1844019 )
[{\tt 1807.10157}].

\bibitem{bordas2009} P. Bordas, V. Bosch-Ramon, J.M. Paredes, \& M. Perucho,
\emph{Non-thermal emission from microquasar/ISM interaction},
\emph{AAP} {\bf 497} (325)
[{\tt 0903.3293}].

\bibitem{drury1983} L.O. Drury,
\emph{An introduction to the theory of diffusive shock acceleration of energetic particles in tenuous plasmas},
\emph{RPP} {\bf 46} (973).

\bibitem{romero2008} G.E. Romero, \& G.S. Vila,
\emph{The proton low-mass microquasar: high-energy emission},
\emph{AAP} {\bf 485} (623)
[{\tt 0804.4606}].

\bibitem{bosch-ramon2006} V. Bosch-Ramon, V., G.E. Romero, \& J.M. Paredes, 
\emph{A broadband leptonic model for gamma-ray emitting microquasars},
\emph{AAP} {\bf 447} (263)
[{\tt astro-ph/0509086}].

\bibitem{reynoso2009} M.M. Reynoso, \& G.E. Romero,
\emph{Magnetic field effects on neutrino production in microquasars},
\emph{AAP} {\bf 493} (1)
[{\tt 0811.1383}].

\bibitem{romero2014} G.E. Romero, \& G.S. Vila,
\emph{Introduction to black hole astrophysics}, Berlin Springer Verlag, 876

\bibitem{vilatesis} G.S. Vila, 
\emph{Ph.D Thesis} 

\bibitem{torres2011} D.F. Torres, \& A. Reimer, A.,
\emph{Hadronic beam models for quasars and microquasars},
\emph{AAP} {\bf 528} (L2)
[{\tt 1102.0851}].

\bibitem{atoyan2003} A.M. Atoyan, \& C.D. Dermer,
\emph{Neutral Beams from Blazar Jets},
\emph{APJ} {\bf 586} (79)
[{\tt astro-ph/0209231}].

\bibitem{kelner2006} S.R. Kelner, F.A. Aharonian, \& V.V. Bugayov,
\emph{Energy spectra of gamma rays, electrons, and neutrinos produced at proton-proton interactions in the very high energy regime},
\emph{PRD} {\bf 74} (034018)
[{\tt astro-ph/0606058}].

\bibitem{mastichiadis2005} A. Mastichiadis, R.J. Protheroe, \& J.G. Kirk,
\emph{Spectral and temporal signatures of ultrarelativistic protons in compact sources. I. Effects of Bethe-Heitler pair production},
\emph{AAP} {\bf 433} (765)
[{\tt astro-ph/0501156}].

\bibitem{aharonian1983} F.A. Aharonian, A.M. Atoyan,  \& A.M. Nagapetyan,
\emph{Photoproduction of Electron-Positron Pairs in Compact X-Ray Sources},
\emph{ASTROPHYSICS} {\bf 19} (187).

\bibitem{takeuchi2010} S. Takeuchi, K. Ohsuga, \& S. Mineshige,
\emph{A Novel Jet Model: Magnetically Collimated, Radiation-Pressure Driven Jet},
\emph{PASJ} {\bf 62} (L43)
[{\tt 1009.0161}].

\end{thebibliography}
\end{document}